 \documentstyle[aas2pp4]{article}

\begin{document}

\title{\bf STOCHASTIC ISOCURVATURE BARYON FLUCTUATIONS,
BARYON DIFFUSION, AND PRIMORDIAL NUCLEOSYNTHESIS}
\author{ H. Kurki-Suonio$^1$, K. Jedamzik$^2$, and G. J. Mathews$^3$}

\affil{$^1$Research Institute for Theoretical Physics and
Department of Physics, University of Helsinki, 00014 Helsinki, Finland}
\affil{$^2$University of California, Lawrence Livermore National Laboratory,
Livermore, CA 94550}
\affil{$^3$University of Notre Dame, Department of Physics, Notre Dame, IN
46556}

\begin{abstract}
We examine effects on primordial nucleosynthesis from
a truly random spatial distribution in the baryon-to-photon ratio ($\eta$). 
We generate stochastic fluctuation spectra characterized by 
different spectral indices and 
root-mean-square fluctuation amplitudes. For the first time we explicitly
calculate the effects of baryon diffusion on the nucleosynthesis yields
of such stochastic fluctuations. We also consider
the collapse instability of large-mass-scale
inhomogeneities.  Our results
are generally applicable to any primordial mechanism
producing fluctuations in $\eta$
which can be characterized by a spectral index.
In particular, these results apply 
to primordial isocurvature baryon fluctuation (PIB) models. 
The amplitudes of scale-invariant
baryon fluctuations are found to be severely constrained by primordial
nucleosynthesis. However, when 
the $\eta$ distribution is characterized by decreasing 
fluctuation amplitudes with increasing length scale,
surprisingly large fluctuation amplitudes on the baryon diffusion scale 
are allowed.
\end{abstract}


\vspace*{-17.0cm}
\hspace*{13.5cm} \mbox{\large HU-TFT-96-20}
\vspace*{16.0cm}

\section{Introduction}

In this paper we investigate the effects of baryon diffusion on the big bang
nucleosynthesis abundance yields for models which are characterized
by a randomly
distributed baryon-to-entropy ratio. 
Such models are distinctively different from the well-studied 
models \markcite{MM93} (Malaney \& Mathews 1993) of 
adiabatic inhomogeneities or baryon
inhomogeneities initially inspired by a first-order QCD phase transition.
We discuss models both with and without the collapse
of a significant fraction of the large-mass scale, high-density regions.
Our study is motivated by the primordial,
isocurvature, baryon (PIB) fluctuation
model of structure formation \markcite{Pee87a,Pee87b} (Peebles 1987ab).
 However, the results and 
constraints derived 
here apply to any theory which produces stochastic fluctuations in the
baryon-to-entropy ratio which can be characterized by a spectral index.

Primordial, isocurvature, baryon fluctuations have 
been  proposed \markcite{Pee87a,Pee87b} (Peebles 1987ab) as a model 
for the formation of cosmic large-scale structure.
In this model
it is envisioned that at early times the universe was filled with
fluctuations in the baryon-to-entropy ratio
characterized by increasing fluctuation
amplitudes with decreasing
mass scale. PIB models with primordial power spectra
of the right normalization
and spectral index ( $-1\ ^<_\sim~n\ ^<_\sim~0$,
where $n=0$ corresponds to white noise)
may be quite successful in reproducing
the statistical properties of the observed
large-scale structure at the present epoch
\markcite{SS92,COP93}(Suginohara \& Suto 1992; Cen, Ostriker, \& Peebles 1993). 
These models have even certain
advantages over other popular structure formation models
(e.g. cold dark matter or mixed dark matter with $n\approx 1$) including
the early formation of structure and the small
peculiar velocity flows on small scales.
An often invoked feature of PIB models is the collapse
of Jeans-mass size ($M_{J}\sim
3\times 10^5M_{\odot}(\Omega_bh^2/0.1)$),
non-linear entropy fluctuations around the
epoch of recombination and the early formation of baryonic,
collisionless dark matter
in the form of black holes or brown dwarfs \markcite{Hog93,Loeb93} 
(Hogan 1993, Loeb 1993).

PIB models, however, are severely constrained by
current observations of the anisotropies in
the cosmic microwave background radiation.
In contrast to cold dark matter or mixed dark matter models,
 PIB models require
an early reionization \markcite{Pee87a, GS89, CSS93, HS94}
(Peebles 1987a; Gorski \&
Silk 1989; Chiba, Sugiyama, \& Suto 1993; Hu \& Sugiyama 1994). 
Moreover, this reionization should only be partial
and occur in the right redshift range. It remains to be shown if this can be
accomplished self-consistently in such models.

PIB models usually assume that the fractional contribution of 
baryons to the critical density
is $\Omega_b\sim 0.1-0.2$ (including remnants, as well as diffuse baryons).
This value is
much larger than that allowed by the light element abundance
constraints of a homogeneous big bang
nucleosynthesis, i.e.  $0.007\ ^<_\sim~\Omega_bh^2\ ^<_\sim~0.023$,
where $h$ is the Hubble parameter in units of 100 km s$^{-1}$ Mpc$^{-1}$ 
\markcite{SKM93,CST95,CF96}
(Smith, Kawano,
\& Malaney 1993; Copi, Schramm, \&
Turner 1995; Cardall \& Fuller 1996). 
This fact,  and the fact that the existence of entropy fluctuations in
general significantly change the abundance yields
during primordial nucleosynthesis,
has led to at least three independent derivations
of nucleosynthesis constraints on the PIB fluctuations  for
small mass scales \markcite{GOR95,JF95,COS95}
(Gnedin, Ostriker, \& Rees 1995; Jedamzik \& Fuller 1995; Copi,
Olive, \& Schramm 1995). On the other hand,
it had already been pointed out in earlier
more schematic studies that the upper limit
on $\Omega_b$ may be substantially larger
that that derived from homogeneous big bang nucleosynthesis when a significant
fraction of high-density regions collapse and \lq\lq swallow\rq\rq their high
$^4{\rm He}$
and $^7{\rm Li}$ yields \markcite{Har68,Zel75,Rees84,SM86}
(Harrison 1968; Zeldovich 1975; Rees 1984; Sale \& Mathews 1986).

Probably the most important result
of all three recent investigations of big bang
nucleosynthesis with stochastic entropy fluctuations
and collapse is that there is no
agreement between calculated primordial abundances
and observationally inferred
abundance limits for any $\Omega_b$ when significant fluctuation power 
is present on mass scales
below the Jeans mass.
This is because high-density fluctuations on
scales smaller than the Jeans mass don't collapse
but rather expand and mix their high
$^4{\rm He}$ and $^7{\rm Li}$ yields with the remaining diffuse baryons 
\markcite{JF94} (Jedamzik \& Fuller 1994).
In other words, a simple extrapolation of the preferred stochastic
PIB fluctuation spectrum to
small mass scales with a spectral index ($-1 < n < 0$)
seems to be ruled out by
nucleosynthesis considerations.

In an attempt to circumvent this constraint, \markcite{GOR95}
Gnedin {\it et al.} 1995 focussed on the
possibility  that entropy fluctuations are strongly phase correlated.  
They considered 
super-Jeans-mass size fluctuations described individually
by a radial power law-
or top hat-  distribution.
Retaining the stochastic nature of
fluctuations \markcite{JF95} Jedamzik \& Fuller 1995 (hereafter; JF95)
speculated on the possibility of
a fluctuation cutoff on mass scales above the Jeans mass.
In both cases the authors
concluded that the total $\Omega_b$ in remnants
and diffuse baryons may be as large as
that reflected by dynamical estimates of the density parameter,
$\Omega\sim 0.1-0.2$
\markcite{Pee86,Kai91,Dek92,Str92}
(Peebles 1986; Kaiser {\it et al.} 1991; Dekel {\it et al.} 1992; 
Strauss {\it et al.} 1992), without being in conflict with nucleosynthesis
abundance constraints.  This is true 
as long as the high-density regions collapse efficiently.
JF95 also pointed out that PIB-like models can produce
intrinsic, large-mass scale
variations in the primordial abundances which may possibly be
inferred from future
observations of $(^2{\rm H/H})$-ratios in high-redshift Lyman-limit clouds
(see also \markcite{JF96} Jedamzik \& Fuller 1996). 

In any case, all previous studies did not adequately address the effects of
dissipative processes during the nucleosynthesis era
on the primordial nucleosynthesis
yields. It is well known that baryon diffusion and hydrodynamic 
expansion during
primordial nucleosynthesis affects the abundance yields when entropy 
fluctuations on
mass scales, $M$, in the range $10^{-22}M_{\odot}\ ^<_\sim~M\ ^<_\sim
10^{-10}M_{\odot}$ are present. It is, perhaps,  
a bold extrapolation to infer the
amplitudes of entropy fluctuations on these small mass scales from 
the amplitudes of
entropy fluctuations on mass scales relevant to
structure formation $(M\gg M_J)$.
Nevertheless, one would expect that
inflation generated spectra should extend over many orders 
of magnitude.  And in any event, disregarding large-scale structure formation,
it is instructive to know which kind of stochastic fluctuation 
spectra (existing with appreciable amplitudes on either only small scales
or both, small and large scales) are compatible with 
nucleosynthesis considerations.

\markcite{GOR95} Gnedin {\it et al.} (1995) 
have attempted to account for dissipation during
nucleosynthesis by using the small-scale, 
inhomogeneous nucleosynthesis yields derived
by \markcite{Mat90} Mathews {\it et al.} (1990). 
The calculations by Mathews {\it et al.}, however,
correspond to the abundance yields produced in a regular lattice
of fluctuation sites,
such as might be produced during a first order QCD transition. 
In contrast, the
few existing baryogenesis scenarios capable of producing PIB-like fluctuations
suggest a stochastic nature of fluctuations \markcite{YS91,DS92}
(Yokoyama \& Suto 1991; Dolgov \& Silk
1992). Moreover, Gnedin {\it et al.} used only those fluctuation parameters
which optimize the upper limit on $\Omega_b$, which is clearly an unrealistic
oversimplification.
JF95 attempted to qualitatively estimate the effects of dissipation during
nucleosynthesis on the abundance yields for 
a random entropy distribution. They
suggested that the $^4{\rm He}$ and $^7{\rm Li}$ yields of a stochastic 
distribution
are generally lowered by the effects of baryon diffusion 
but should still be higher
than the corresponding yields in a homogeneous big 
bang nucleosynthesis scenario at
the same $\Omega_b$.

In this paper we compute primordial nucleosynthesis yields for random entropy
fluctuations with different spectral indices and fluctuation amplitudes. 
We explicitly incorporate effects of baryon diffusion during the
nucleosynthesis epoch.  Our
calculations serve to place constraints 
on the homogeneity in the baryon-to-photon
ratio independent of any gravitational collapse. 
We also combine our results with
PIB-like models of structure formation in order to 
derive approximate constraints on
the allowed isocurvature power spectrum. In Section 2 
we give the relevant physical scales entering the problem,
introduce our models of stochastic baryon inhomogeneity
which are based on the models of JF95,
and explain our numerical
techniques for the calculation of primordial abundance yields.
In Section 3 we present our results and in Section 4 we
draw conclusions.  

\section{Random Baryon Fluctuations and Primordial Nucleosynthesis}

\subsection{Physical Length and Mass Scales}

In what follows we will give all length scales in comoving units
at the present epoch. Often we will associate a length scale, $\lambda$ ,
with a baryonic mass scale, $M_b(\lambda )$, by employing

\begin{equation}
M_b={\pi\over 6}\bar{\rho_b}\lambda^3\ ,\label{eq:1}
\end{equation}
where $\bar{\rho_b}$ is the average baryon density at the present epoch.
There are three relevant physical scales in our problem. 
First, there is the
neutron diffusion length scale, $d_n(T\approx 500{\rm keV)}$ , determined at 
the epoch of weak
freeze-out in the early universe. This scale divides
length scales, $L$ , 
which have been
homogenized by baryon diffusion before the onset of primordial nucleosynthesis,
$L\ ^<_\sim\ d_n$, from length scales on which inhomogeneities in 
the baryon-to-photon ratio $\eta$ can persist
through the beginning of primordial nucleosynthesis, $L\ ^>_\sim\ d_n$,

\begin{equation}
d_n(T\approx 500 {\rm keV})\approx 4\times 10^{-5}{\rm pc}\ .\label{eq:2}
\end{equation}
In deriving Eq.~\ref{eq:2} we assume $\Omega_bh^2\ ^<_\sim\ 1$ locally
(see \markcite{JF94} Jedamzik \& Fuller 1994). 
Using Eq.~\ref{eq:1} with $d_n=\lambda$,
we find the corresponding mass scale 

\begin{equation}
M^d_{\rm
500keV}\approx 1\times
10^{-22}M_{\odot}(\Omega_bh^2/0.0125)\ .\label{eq:3}
\end{equation}

The second length scale is the neutron diffusion scale at the approximate 
completion
of freeze-out from nuclear statistical equilibrium

\begin{equation}
d_n(T\approx 10{\rm keV})\approx
8\times 10^{-2}{\rm pc}(\Omega_bh^2/0.0125)^{-{1 \over 2}}
\ .\label{eq:4}
\end{equation}
This scale corresponds to a baryonic mass scale
\begin{equation}
M^d_{\rm
10keV}\approx 1\times
10^{-12}M_{\odot}(\Omega_bh^2/0.0125)^{-{1\over 2}}\ .\label{eq:5}
\end{equation} 
We note here that the relevant $M^d$ at the completion
of the freeze-out process
is also sensitive to  the local $\Omega_b h^2$ because
the termination temperature of the primordial nucleosynthesis
process is itself a function of $\Omega_b h^2$. 
Small local $\Omega_b h^2$ imply
relatively large $M^d\gg10^{-12}M_{\odot}$.
Regions separated by length scales, $L>d^n_{\rm 10keV}$, 
are not in contact through
baryon diffusion during the epoch of primordial nucleosynthesis. 
If any one region of
the universe is homogeneous on length scales $d^n_{\rm 10keV}$, 
but inhomogeneous on
larger scales, the nucleosynthesis yields can be simply computed by 
an appropriate abundance average
over different homogeneous big bang nucleosynthesis scenarios 
at different $\eta$.
This is what has been done in all prior studies \markcite{GOR95, JF95, CST95} 
(Gnedin {\it et al.} 1995; JF95;
Copi {\it et al.} 1995). However, 
when small-scale inhomogeneities are present a
reliable calculation has to at least cover the length scale range from 
$d^n_{\rm 500keV}$ to
$d^n_{\rm 10 keV}$. Our calculations achieve this.
Finally, there is the Jeans length

\begin{equation}
\lambda_J\approx 5\times 10^{-2}{\rm Mpc}
\biggl({\Omega_bh^2\over 0.1}\biggr)^{-{1\over
2}}\ ,
\end{equation}
yielding the Jeans mass

\begin{equation} 
M_J^b\approx 3\times
10^5M_{\odot}(\Omega_bh^2/0.1)^{-{1\over 2}}\ .\label{eq:7}
\end{equation}
The ultimate fate of overdense regions with $M\ ^>_\sim~M_J^b$ is collapse, 
whereas isolated, overdense regions with $M\ ^<_\sim~M_J^b$ 
will expand and mix their
nucleosynthesis yields with the remaining diffuse baryons. 
It should be noted that
$M_J^b$ is {\it not} the true Jeans mass before the epoch of recombination. 
It rather
compares the effects of baryonic pressure with baryonic 
self-gravity and determines if
a fluctuation will expand or contract. 
It coincides with the post-recombination Jeans
mass when all photon stresses can be neglected, 
and we loosely refer to it here as the Jeans mass.

\subsection{Models of Primordial Isocurvature Baryon Fluctuations}

PIB-like models may be described by the root-mean-square (rms) 
of the fractional variation
in primordial baryon density as a function of mass scale $M$,
\begin{equation}
\biggl({\delta\rho_b\over\rho_b}\biggr)_{\rm rms}(M)=
\biggl({\delta\rho_b\over\rho_b}\biggr)_{\rm rms}(M_N)
\biggl({M\over
M_N}\biggr)^{-{1\over 2}-{n\over 6}}\ .\label{eq:8}
\end{equation} 
Here $M_N$ is an abritrary normalization mass scale,
$(\delta\rho_b/\rho_b)(M_N)$ is the fluctuation rms on $M_N$,
and $n$ is a spectral index. 
For the primordial abundance calculations, however, the
knowledge of the statistics of the distribution is not sufficient. 
Rather, we need to
specify a distribution in baryon density as a function of spatial coordinate,
$\rho_b(x)$. It is not a unique procedure to infer realizations of 
a distribution in
baryon-to-entropy from  Eq.~\ref{eq:8}. We follow JF95 and consider a Gaussian 
random variable,
$A(x)$. We generate stochastic baryon-to-photon distributions, 
$\eta (x)$, by the relations
\begin{equation}
\eta (x)=\eta_NA^2(x)\ ,\label{eq:9}
\end{equation}
or
\begin{equation}
\eta (x)=\eta_NA^{10}(x)\  .\label{eq:10}
\end{equation}
Here $\eta_N$ is a normalization constant. Note that the $\eta (x)$ in 
Eq.~\ref{eq:9} and Eq.~\ref{eq:10} are
chosen positive definite to avoid the introduction of antimatter domains. 
It has been shown in JF95 that the resultant distribution in 
$\eta$ is highly non-Gaussian on
small scales, in particular the transformations in 
Eq.~\ref{eq:9} and Eq.~\ref{eq:10} introduce phase correlations
between the different Fourier modes. However, the distribution quickly 
approaches Gaussian character
on larger scales.

We will employ a fluctuation cutoff length scale, $\lambda_c$ , 
below which any regions are assumed to be homogeneous. 
As long as we choose $\lambda_c\ ^<_{\sim}\ d^n_{\rm 500 keV}$, however, there
will be no physical relevance to this scale
since baryon diffusion prior to the nucleosynthesis epoch
will provide a natural cutoff scale. Aside from any 
non-Gaussian character of the
distribution, the statistics of the distribution is then 
fully described by three
quantities. These are the effective spectral index $n$ in Eq.~\ref{eq:8}, 
the rms fluctuation
on the cutoff scale, $(\delta\rho_b /\rho_b)_{\lambda_c}$, 
and the magnitude of the cutoff scale, $\lambda_c$. 

\subsubsection{Models without Collapse}

We will study three models of small-scale stochastic baryon inhomogeneity with
fluctuations in $\eta$ spanning a range from very small scales $\sim d^n_{\rm
500keV}$ to quite large scales well above $d^n_{\rm 10keV}$.
We refer to these
models as Model 1, 2, and 3.  They are characterized by
\begin{equation}
{\rm Model\ 1:}\ n\approx -3,\
\biggl({\delta\rho_b\over\rho_b}\biggr)(\lambda_c)=\sqrt{2}\ ,\label{eq:11} 
\end{equation}
\begin{equation}
{\rm Model\ 2:}\ n\approx -3,\
\biggl({\delta\rho_b\over\rho_b}\biggr)(\lambda_c)\approx 27\ ,\label{eq:12}  
\end{equation}
\begin{equation}
{\rm Model\ 3:}\ n\approx -1.5,\
\biggl({\delta\rho_b\over\rho_b}\biggr)(\lambda_c)\approx 27\ .\label{eq:13} 
\end{equation}
In all three models we employ a cutoff-scale $\lambda_c=6.8\times 10^{-5}$pc,
which is only slightly larger than $d^n_{\rm 500 keV}$.
Model 1 is generated by employing Eq.~\ref{eq:9} 
and Models 2 and 3 are generated by
employing Eq.~\ref{eq:10}. It is seen 
from Eq.~\ref{eq:8} that Models 1 and 2 are chosen to have a close to
scale invariant spectrum, $(\delta\rho_b /\rho_b)\approx const$. 
Note that these models do
not incorporate any aspects of structure formation such as the collapse of
high-density, super-Jeans-mass size regions. 
The specific realizations of the distributions characterized by 
Eq.~\ref{eq:11} - Eq.~\ref{eq:13}
which are used to calculate big bang nucleosynthesis
abundance yields for Models 1-3 are shown in the top panels of
Figures 1-3. These panels show the spatially varying baryon-to-photon 
ratio divided by the average baryon-to-photon ratio as a function 
of spatial coordinate. In order to
produce these distributions we generated $2\times 10^4$ Fourier modes. 
Our simulation therefore has a dynamic length scale range of $2\times 10^4$. 

It is evident that we
examine a one-dimensional planar analogue to the full 
three-dimensional theory. 
This is a necessity in order to have sufficient length scale coverage 
to confidently calculate
primordial abundance yields when baryon diffusion is occurring. Our results are
therefore only an approximation to a three-dimensional distribution in
$\eta$. Even though abundance yields in inhomogeneous nucleosynthesis
have been found to be geometry dependent \markcite{Mat90} (Mathews
et al. 1990), we expect our one-dimensional
calculations to uncover the trends and magnitudes
of the effects of baryon diffusion on the nucleosynthesis yields. 

There is an additional uncertainty regarding 
the generality of our results for the
abundance yields computed from the specific distributions shown in Figures 1-3.
Different distributions than those shown in Figures 1-3, albeit with the same
statistical properties, may well give different results for 
the primordial abundance
yields. We, however, expect this form of \lq\lq simulation variance\rq\rq\ 
(similar to the variance in simulations of the formation
of large-scale structure) to be
small in magnitude. Nevertheless, it should be kept in mind that 
particularly for the scale invariant spectra Model 1 and 2
there is a small amount of
simulation variance.

\subsubsection{Models with Collapse}

In the second part of our study we will combine Models 1-3 
for small-scale stochastic baryon inhomogeneity with large-scale PIB-like
fluctuations and the collapse of overdense, super-Jeans mass size regions. 
We will refer to these models as Models 1C - 3C. 
We follow JF95 and exclude the abundance
yields of regions from the abundance average, whenever 
the fractional variations in
baryon density exceeds a critical value, 
$(\delta\rho /\rho)\geq \Delta_{cr}$, and
when the region's mass scale satisfies, 
$M>M_J^b$. As argued in JF95, for slightly
non-linear, primordial fluctuations, $\Delta_{cr}\ ^>_{\sim}\ 1$, 
we expect early collapse
and, possibly, the formation of dark remnants.

Similarly to the spectrum of small-scale inhomogeneities, and as done 
in JF95, we will
have to introduce a fluctuation cutoff mass scale for the large-scale PIB
fluctuations. This is due to the impossibility of generating
 Fourier modes covering the
whole mass scale range from the very small mass scales, 
$M\sim 10^{-22}M_{\odot}$, to
the mass scales relevant for structure formation, 
$M\ \gg 3\times 10^5M_{\odot}$.
In fact, in order to calculate the abundance yields 
for PIB spectra with small- and
large-scale power, we will only slightly modify 
the procedure of JF95. Whereas JF95
assumed any region with mass $M<M_J^b$ to be homogeneous, 
we will assume that any such
region's abundances are well 
approximated by the abundance yields computed from the
small-scale inhomogeneous distribution shown in the top panels of Figure 1-3.

It is common practice in the field of structure formation to 
describe the behavior of
fluctuation amplitudes on length (mass) scale by a 
primordial power spectrum, $P(k)$.
In the absence of any significant phase correlations the rms 
fluctuation amplitudes in
the fractional variations of primordial baryon density as a function of mass 
scale can be derived from the power spectrum by
\begin{equation}
\biggl({\delta\rho_b\over\rho_b}\biggr)_{\rm rms}^2(M)
 =\int_0^{k(M)}d{\rm ln}k\ k^3P(k)\ .
\end{equation}
In this expression we derive the relationship for $k(M)$ from 
Eq.~\ref{eq:1} with
$\lambda=(2\pi /k)$, corresponding to sharp $k$-space filtering. 
The approximate primordial fluctuation
power per unit logarithmic wave vector interval (e.g. $k^{3/2}P^{1/2}(k)$)
which characterizes our models of inhomogeneous 
nucleosynthesis and the collapse of
overdense regions are shown in Figure 4. 
Note that the figure shows the assumed truly
primordial and unprocessed power spectra. In particular, 
any microphysical processing
and, for example, 
different gravitational growth factors for different Fourier modes,
usually accounted for by a transfer function, are not included.

The behavior of these models on the nucleosynthesis scale
is shown on the left hand (large $k$) side of Figure 4, whereas
the behavior of the models on the large scales relevant for collapse is shown
on the right hand (small $k$) side of Figure 4.
The cutoff for each model is shown on the right.
In this figure it can be 
seen that the extended mass scale range in our problem forced us to resort to
rather contrived power spectra. In particular, all three models do not have any
fluctuation power on length scales roughly in 
the range $1{\rm pc}\ ^<_\sim~\lambda\
 ^<_\sim\ 1{\rm Mpc}$. These power spectra, however, do
approximate a scenario with a white noise $(n=0)$ large-scale tail, 
turning over to a
scale-invariant spectrum $(n=-3)$ slightly above the Jeans-mass 
scale, and either staying
scale-invariant through the scales relevant 
for diffusion during nucleosynthesis
(Model 1C, 2C), or having a second turnover 
on these scales approaching spectral index
$n=-1.5$ (Model 3C).  

Specifically, Models 1C-3C are characterized by the following parameters:
\begin{eqnarray}
{\rm Model\ 1C:}&  n_s\approx -3;\
\biggl({\delta\rho_b\over\rho_b}\biggr)(\lambda_{c}^s)
 =\sqrt{2}\times\sqrt{2};\nonumber \\
&\lambda_{c}^s=6.8\times 10^{-5}{\rm pc}\ , \\
& n_l\approx 0;~
\biggl({\delta\rho_b\over\rho_b}\biggr)(\lambda_{c}^l)=\sqrt{2};\nonumber \\
&\lambda_{c}^l=6.8\times 10^{-1} h^{-1} {\rm Mpc};\
\Delta_{cr}=1.5\ ,\nonumber
\end{eqnarray}
\begin{eqnarray}
{\rm Model\ 2C:}&    n_s\approx -3;\ 
\biggl({\delta\rho_b\over\rho_b}\biggr)(\lambda_{c}^s)
 =\sqrt{2}\times 27;\nonumber \\
&\lambda_{c}^s=6.8\times 10^{-5}{\rm pc}\ , \\
& n_l\approx 0;\ \biggl({\delta\rho_b\over\rho_b}\biggr)(\lambda_{c}^l)
 =27;\nonumber \\
&\lambda_{c}^l=9.5\times 10^{-2} h^{-1} {\rm Mpc};\ \Delta_{cr}=2\ ,\nonumber 
\end{eqnarray}
\begin{eqnarray}
{\rm Model\ 3C:}& n_s\approx -1.5;\
\biggl({\delta\rho_b\over\rho_b}\biggr)(\lambda_{c}^s)
 =\sqrt{2}\times 27;\nonumber \\
&\lambda_{c}^s=6.8\times 10^{-5}{\rm pc}\ ,\\
& n_l\approx 0;\ \biggl({\delta\rho_b\over\rho_b}\biggr)(\lambda_{c}^l)
 =27;\nonumber \\
&\lambda_{c}^l=9.5\times 10^{-2} h^{-1} {\rm Mpc};\ \Delta_{cr}=2\ . \nonumber 
\end{eqnarray}
Here the upper lines for each model give the parameters for the small-scale
fluctuations, whereas the lower lines give the parameters for the large-scale
fluctuations. 
Note that the small-scale distribution of Model 1C (2C, 3C) is identical
to the small-scale distribution of Model 1 (2, 3). 
The factors of $\sqrt{2}$ in the
$(\delta\rho /\rho) (\lambda_{c}^s)$ have only been written to illustrate 
that the large-scale fluctuation rms has to be added 
to the small-scale fluctuation rms in
quadrature. The cutoff scales $\lambda_{c}^l$ and fluctuation
amplitudes $(\delta\rho_b/\rho_b)(\lambda_c^l)$ 
have been chosen such that the PIB
fluctuation parameters are roughly normalized to yield an rms in the fractional
variation of baryon density on the scale $8h^{-1} {\rm Mpc}$ 
of unity at the present
epoch (JF95). Here we assumed full reionization. 
It is evident that the chosen cutoff scales for the 
large-scale PIB-fluctuations satisfy $\lambda_c^l\ >\ \lambda_J$.

\subsection{Numerical Techniques for the Calculation
of Primordial Abundance Yields}

The small-scale inhomogeneous nucleosynthesis calculations made use of the
nucleosynthesis code developed by \markcite{KM89,Kur90,KM90,Kur92} 
Kurki-Suonio {\it et al.} (1989, 1990a, 1990b,
1992).  The nuclear reaction rates were updated according to Kawano
\markcite{KSM93} {\it et al.} (1993).
Regarding weak reaction rates, we used the \markcite{Wag73}
Wagoner (1973) polynomial for the
weak reaction rates, with $\tau_n = 887.0$s.  We included all recently
determined corrections to this rate as relevant
\markcite{Dic82,KK94,Sec93,Ker93} (Dicus {\it et al.} 1982;
Kernan \& Krauss 1994; Seckel 1993; Kernan 1993). We also corrected for
effects of finite spatial resolution and time step effects. The net correction
factor to $Y_p$ from all of these effects was only -0.0012.
(Note that the correction +.0017 found by \markcite{Ker93} Kernan (1993)
 by using a
shorter time-step does not apply to our code since the code used here is
independent from the one used by Kernan.)
We use all the relevant diffusion coefficients for protons and neutrons 
as given by
\markcite{JF94} Jedamzik \& Fuller (1994).
However, we do not incorporate the effects of late-time hydrodynamic expansion
and neutrino cooling of inhomogeneities into our calculations
\markcite{JFM94} (Jedamzik, Fuller, \& Mathews 1994) 
which renders our calculated
$({}^7{\rm Li/H})$ ratios somewhat higher than when hydrodynamic expansion
is included.

A key challenge in the present work was the necessity to utilize
a very large spatial grid (40 000 zones).   It was thus
essential to optimize the speed of computations.  
We therefore included only the 8 most important light
isotopes in our reaction network.   The reactions leading to isotopes heavier
than A$=7$ were included as sinks in the reaction flow.  Since there is little
flow from heavier isotopes back to lighter, this is an excellent approximation
for the present application.  
The most time consuming part of the nucleosynthesis
calculation is the inversion of an
$8\times8$ reaction matrix which must be 
performed at each zone at each time-step.  Since
this inversion does not easily vectorize, we found it best to 
run on a fast scalar machine.

\section{Results}

In this section we present the results of our inhomogeneous primordial
nucleosynthesis calculations.
Figures 1-3 show the employed random distributions 
in $\eta(x)/\langle \eta \rangle$
for the small-scale baryon inhomogeneities of Models 1-3 (and Models 1C-3C)
in the top panels.
Nucleosynthesis yields for the various isotopes and models 
are given along the side panels of Figures 1-3. 
The three panels on the left-hand-side of the figures
give the $^4$He mass fraction, $Y_p$ ,
$({}^2{\rm H}/^1{\rm H})$ -number ratio, 
and $({}^7{\rm Li}/{\rm H})$ -number ratio
as a function of average baryon-to-photon ratio $\eta$
for the models without collapse. The three panels on the right-hand-side
of the figures give the abundance yields for the same light elements 
as a function of baryon-to-photon ratio for
the models with collapse.
For reference the fractional contribution of baryons to the critical density,
$\Omega_b$ , can be obtained from the baryon-to-photon ratio $\eta$ by
$\Omega_b=0.0112(\eta /3\times 10^{-10})h^{-2}$.

\subsection{Models without collapse}

For the panels on the left-hand-side the three lines
correspond to: Solid lines--the results of a homogeneous
big bang nucleosynthesis (hereafter; BBN) calculation at the same average
$\eta$; Dotted lines-- the results of an inhomogeneous BBN calculation of the
distribution shown in the top panel of the figure but {\it without} taking into
account of the effects of baryon diffusion; and  Dashed lines-- the results
of our inhomogeneous BBN calculations which include the
effects of baryon diffusion.

In the left-hand-side panels of
Figure 1 we show nucleosynthesis yields for Model 1. 
It is seen that the abundance yields of the inhomogeneous BBN Model 1 are not
very different from the yields of a homogeneous BBN scenario at the same
average $\eta$ (with the possible exception of $^7$Li).
This is mainly because the fluctuation amplitudes in Model 1 are 
relatively small.
The trends in Model 1 can be best understood when compared to the results of
Model 2 which are shown in the left-hand-panels of Figure 2. Note that
both Model 1 and 2 have an approximately scale-invariant 
spectrum ($n\approx -3$) but
that Model 2 has much larger fluctuation amplitudes than Model 1.
It is evident, particularly from the results of Model 2, that for close to
scale-invariant spectra baryon diffusion does {\it not} 
significantly change the
calculated abundance yields. This can be inferred by comparing the calculated
abundance yields of the inhomogeneous distribution where baryon diffusion 
is accounted
for (dashed line) to those where baryon diffusion is neglected (dotted line).
This trend is also observed in Model 1 (Figure 1). That baryon diffusion
is comparatively unimportant can be understood by inspection of the baryon
distributions in Figure 1 and 2. The bulk of the baryon number in both models
is included in fairly extended high-density regions with only relatively small
variations in $\eta$ around the approximate average baryon-to-photon ratio of
these extended regions. It is known that the effects of baryon diffusion are
maximized when significant fluctuations exist on scales 
comparable to the neutron
diffusion scale during the epoch of primordial nucleosynthesis
\markcite{AHS87} (Applegate, Hogan \& Scherrer 1987). The effects of
baryon diffusion on Models 1 and 2 are small since the typical length scale
of extended high-density regions is much larger than the neutron diffusion
length scale.

For most $\eta$ values the $^7$Li yields are substantially higher
for the inhomogeneous distribution than the corresponding $^7$Li yields in a
homogeneous BBN scenario.  
This is a generic feature of inhomogeneous primordial
nucleosynthesis \markcite{AFM87} (Alcock et al. 1987). 
Because of this, inhomogeneous BBN scenarios are only compatible
with observations if there is moderate to significant depletion of $^7$Li
in low-metallicity Pop II halo stars. In fact, there may be evidence
for $^7$Li depletion in Pop II halo stars \markcite{PDD92}
(Pinsonneault et al. 1992). 
However, the situation
remains unresolved \markcite{SM96} (Schramm \& Mathews 1996). 

If one allows for the possibility of $^7$Li depletion, and 
assumes that the primordial $({}^2{\rm H}/{}^1{\rm H})$ -ratio
is as large as may have been
inferred from some (but not all) observations of 
deuterium in Lyman-limit clouds
at high redshift \markcite{Son94,RH96,TFB96,BT96}
(Songailia et al. 1994; Rugers \& Hogan 1996; but see also 
Tytler, Fan, \& Burles 1996; Burles \& Tytler 1996), 
Model 1 is marginally
consistent with observations for some range in $\eta$. 
In contrast, Model 2 has no range in $\eta$
consistent with the observationally inferred light-element abundances.
We conclude that the fluctuation amplitudes of stochastic, approximately
scale-invariant baryon inhomogeneities, are severely constrained 
by the possible
overproduction of $^4$He and/or $^7$Li 
during the epoch of big bang nucleosynthesis
even when the effects of baryon diffusion are included.

In Model 3 we assumed a fluctuation spectrum with 
spectral index $n\approx -1.5$ and
with significant fluctuation amplitudes. 
In contrast to Models 1 and 2 this model
is characterized by decreasing fluctuation amplitudes with 
increasing length scale as
evident from the $\eta$ distribution shown in
the top panel of Figure 3. In a similar fashion to Figures 1 and 2 the
left-hand-side panels of Figure 3 show the abundance yields of Model 3. 

The results of Model 3 are, perhaps,  somewhat startling. 
It is evident from comparison
of the dotted and dashed lines in Figure 4 that 
for the parameters employed in Model 3
baryon diffusion {\it significantly} changes the 
abundance yields. In particular,
the $^4$He mass fraction in Model 3 is very close to that of a homogeneous BBN
scenario and much smaller than the calculated $^4$He mass fraction for the
inhomogeneous distribution when baryon diffusion is neglected. 
For most of the range
in $\eta$ $^7$Li yields are substantially smaller and deuterium yields are 
larger than those in inhomogeneous BBN scenarios without diffusion.

The effects of baryon diffusion are important since significant fluctuation 
amplitudes
exist on the scale of the neutron diffusion length during the epoch of
primordial nucleosynthesis. Fluctuations on larger spatial scales, 
which are much
less affected by neutron diffusion and which may grossly 
overproduce $^4$He and $^7$Li,
are substantially smaller than those on the neutron diffusion length due to the
large spectral index of the distribution. In  other words, the average $\eta$
in extended regions is close to the average 
baryon-to-photon ratio for the whole distribution. 

Assuming high $^2$H and significant $^7$Li depletion in Pop II
halo stars there is some range for $\eta$ in Model 3 
which can agree with all the
observationally inferred primordial abundance yields. This is quite surprising
since it implies that random variations in the 
baryon-to-entropy with rms amplitude as
large as $(\delta\rho_b/\rho_b)\sim 30$ on $d^n_{\rm 500 keV}$ 
are not ruled out by the observations of $^2$H, $^4$He, 
and $^7$Li in low-metallicity environments.
Nevertheless, if in future it is shown convincingly that either the primordial
$({}^2{\rm H}/{}^1{\rm H})_p\approx 3\times 10^{-5}$ is low or the primordial
$({}^7{\rm Li}/{\rm H})_p\approx 2\times 10^{-10}$ is low, then significant
baryon-to-entropy fluctuations 
during the epoch of primordial nucleosynthesis are
probably ruled out. This applies for both, 
inhomogeneities in a nearly regular lattice
(such as might be generated by a first-order QCD phase transition), 
or stochastic
inhomogeneities of arbitrary statistical properties as may be generated 
during inflation.

\subsection{Models with collapse}

We now turn our attention to the results of Models 1C, 2C, 
and 3C which incorporate
the collapse of overdense, large-scale regions. The abundance yields
from such overdense, super-Jeans-mass size regions will be excluded
from the abundance average since these regions either may form 
black holes, brown
dwarfs, and/or stellar remnants,  or might undergo sufficient
early stellar processing and chemical
evolution which would render the regions 
unsuitable for the determination of primordial abundances (JF95).
The results for the primordial abundances of Model 1C, 2C, 
and 3C are presented in the
right-hand-side columns of Figure 1, 2, and 3, respectively.
Each panel in these figures shows the results for: Dotted lines -- 
a model which does not
have any sub-Jeans-mass size fluctuation power as in JF95; Dashed lines --
a model which includes fluctuations on scales below the Jeans length
and explicitly treats diffusion as specified in Section 2;
and  for comparison, Solid lines -- the results of a standard homogeneous 
BBN calculation.

As already discussed in detail in JF95, models which have a fluctuation cutoff
above the Jeans mass, and which have high efficiencies for collapse of 
overdense regions,
generally yield substantially lower $^4$He mass fractions $Y_p$, higher
$({}^2{\rm H}/{}^1{\rm H})$ -number ratios, and comparable to higher
$({}^7{\rm Li/H})$ -number ratios than a homogeneous BBN scenario.
The modifications to these results by the introduction of small-scale power
on scales below the Jeans scale studied here are evident from the
dashed lines. In general, small-scale fluctuations enhance $^4$He and $^7$Li
yields over those yields from PIB models without small-scale power. This is
easily understood since regions fluctuating in $\eta$ 
mostly yield larger $Y_p$ and
$({}^7{\rm Li/H})$ ratios (even when baryon diffusion is included) than regions
which are homogeneous in $\eta$ (as apparent from the left-hand-side panels of
Figures 1-3). 

If PIB models with collapse and small-scale fluctuations even exceed the
$^4$He and $^7$Li yields of a standard homogeneous BBN scenario largely depends
on rms fluctuation amplitude. 
Whereas Model 1C is marginally compatible
with observationally inferred primordial abundances Models 2C and 3C, which
have large rms amplitudes in $\eta$, 
either overproduce $^7$Li and/or $^4$He for
large $\eta$ or overproduce deuterium for small $\eta$. 
It is evident that none of the models which include
small-scale fluctuations can agree with observations for large
average baryon-to-photon ratios ($\eta\ {}^>_\sim\ 3\times
10^{-10})$ mainly due to an overproduction of $^7$Li. This is not the case for
PIB models
which are homogeneous on scales below the Jeans scale (JF95).
This implies that the
introduction of small-scale fluctuations into stochastic PIB models 
considerably
narrows, or even excludes, the ranges of $\eta$ which may have agreed with
observationally inferred primordial abundances for models without small-scale
fluctuations.

\section{Conclusions}

We have investigated the primordial nucleosynthesis process for models
which include random variations in the baryon-to-photon ratio.
Our study explicitly accounts for the effects of baryon diffusion.
Analyzing three different stochastic distributions in $\eta$, which
have different fluctuation amplitudes and spectral indices, we
have found that the statistical
properties of stochastic baryon-to-photon inhomogeneities are
severely constrained by the possible overproduction of either
$^7$Li and/or $^4$He for large average $\eta$ or 
deuterium for small average $\eta$.
In particular, we have found that the rms amplitudes of scale-invariant
$\eta$ distributions, characterized by almost equal fluctuation
amplitudes on all scales, are constrained to be not larger than a few.

In contrast, the constraints on the rms amplitudes of
distributions characterized by decreasing fluctuation 
amplitudes with increasing
mass scale are much weaker.
In such models rms amplitudes as large as $\sim 30$ 
on the neutron diffusion scale at weak freeze-out are currently not ruled
out by observationally inferred primordial abundance limits. 
Nevertheless, these
constraints may be considerably tightened if it is ever 
convincingly shown in the future that either
depletion of primordial $^7$Li 
in low-metallicity halo stars is negligible or that 
the primordial deuterium-to-hydrogen
ratio is low $({}^2{\rm H}/{}^1{\rm H})\sim 3\times 10^{-5}$. 

In the second part of our study we have considered stochastic baryon
inhomogeneity extending from the very small baryon diffusion scale
during the epoch of primordial nucleosynthesis to very large scales
above the post-recombination Jeans length, and the early collapse of
overdense super-Jeans-mass size regions. 
In agreement with JF95 we have found that
significant fluctuation amplitudes on scales below the Jeans scale
considerably narrow (or exclude) the parameter space of such PIB-like models
which may agree with observationally inferred primordial abundance
limits otherwise. We conclude that PIB fluctuations may only
be consistent with observationally inferred primordial abundance limits
for large $\Omega_b$ if either fluctuations are highly phase correlated
(Gnedin {\it et. al.} 1995) or if there exists a fluctuation cutoff on scales
above the Jeans scale (JF95).

\acknowledgments

This work was performed in part under the auspices of
the US Department of Energy by the Lawrence Livermore National
Laboratory under contract number W-7405-ENG-48.
Work at University of Notre Dame was
supported in part by DOE Nuclear Theory grant DE-FG02-95ER40934.
Computation was done in part at the Center for Scientific
Computing (Finland).

\section*{Figure Captions}
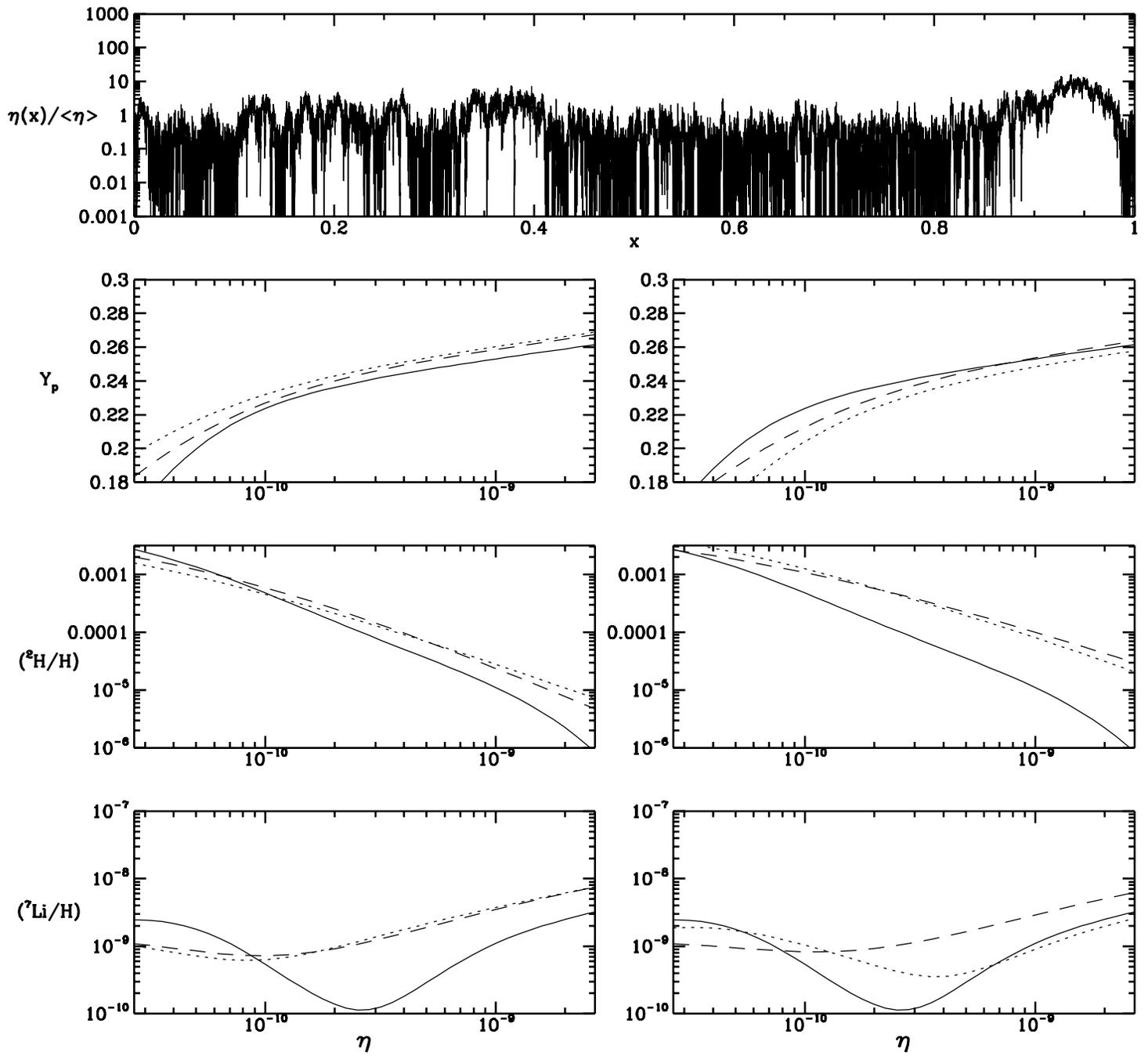
\begin{figure}
\caption{Baryon-to-photon distribution(top panel) and nucleosynthesis
yields for Model 1 (left)  and Model 1C (right side panels). 
A value of unity on the ordinate of the top panel corresponds 
to $2\times 10^4\lambda_c$ 
for Model 1 (or $2\times 10^4\lambda_c^s$ for Model 1C),
where $\lambda_c$ ($\lambda_c^s$) is specified in the text. 
The side panels give the calculated
$^4$He mass fraction $Y_p$, ($^2$H/$^1$H) -number ratio, and ($^7$Li/H) -number
ratio as a function of the average $\eta$.
The left-hand side panels give results for the model without collapse 
(Model 1).
The dashed lines are for
the inhomogeneous distribution when baryon diffusion is included.
The dotted lines are for the inhomogeneous distribution when baryon diffusion
is neglected.  For comparison, the solid lines are abundance yields for a 
homogeneous BBN calculation.
The panels on the right-hand-side show the abundance
yields for models with collapse of high density regions (Model 1C).
The dashed lines are for an inhomogeneous distribution with collapse
when small-scale fluctuation as shown in the top panel are present.
The dotted lines are for an inhomogeneous distribution with collapse when 
there are no small-scale fluctuations.  
The solid lines are for a homogeneous BBN calculation.
Note that the
large-scale fluctuation part of Model 1C is not shown in this figure.}
\label{fig1}
\end{figure}

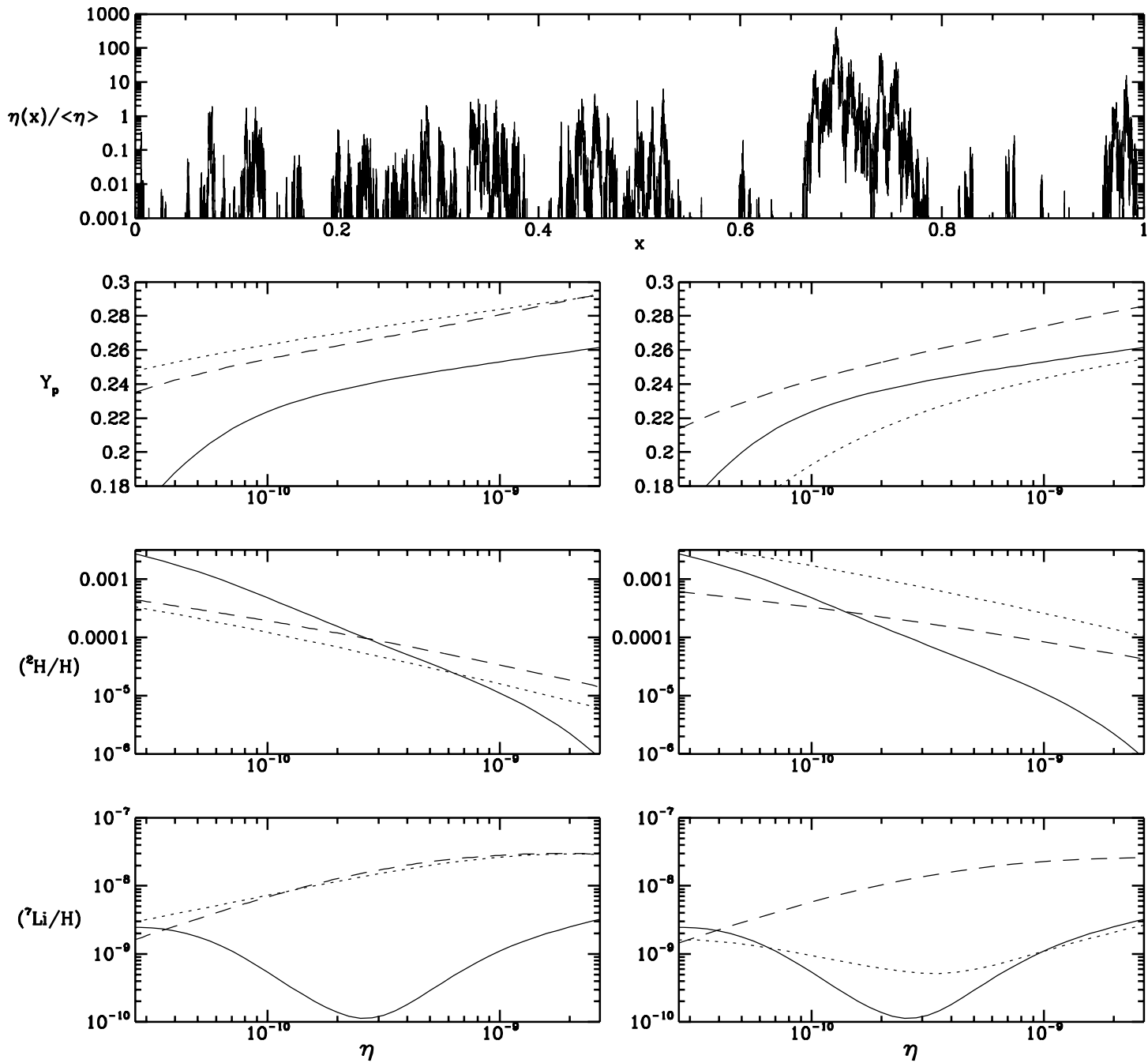
\begin{figure}
\caption{Same as Figure 1, but for Model 2 and Model 2C.}
\label{fig2}
\end{figure}

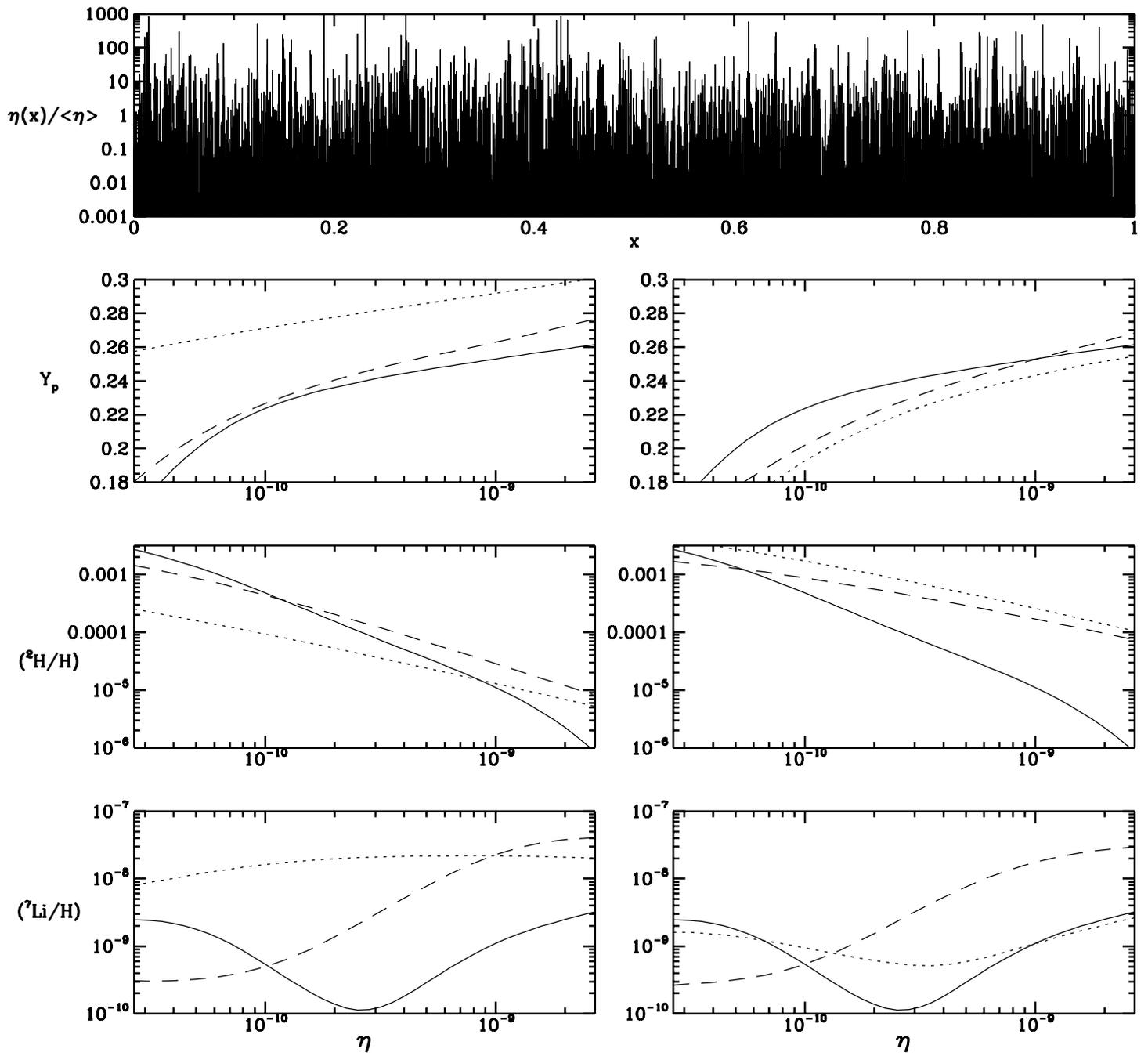
\begin{figure}
\caption{Same as Figure 1, but for Model 3 and Model 3C.}
\label{fig3}
\end{figure}

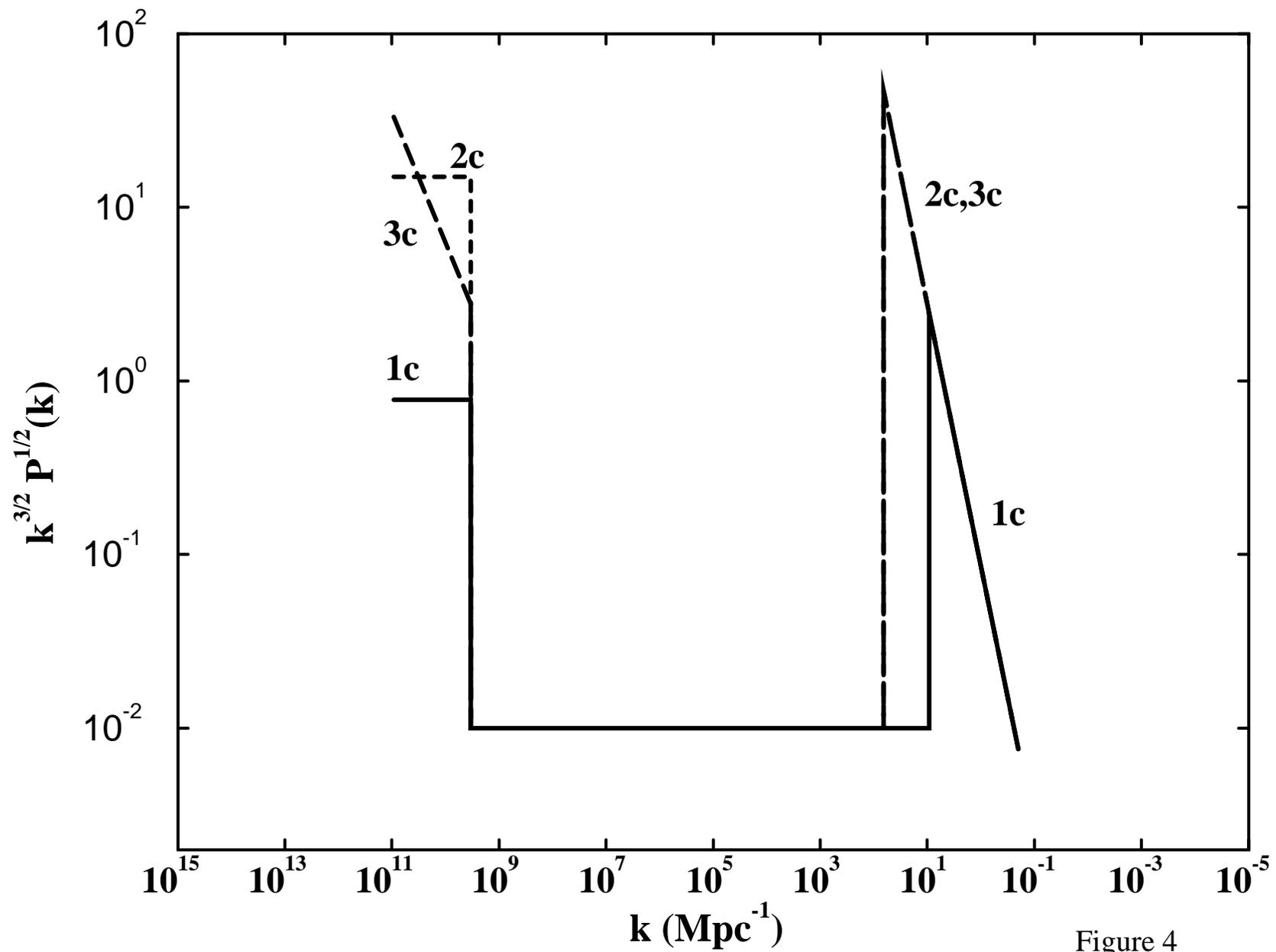
\begin{figure}
\caption{Illustration of the primordial
fluctuation power per unit logarithmic wave vector
interval $P^{1\over 2}(k)k^{3\over 2}$ as a function of wave vector $k$
for the models employed in the nucleosynthesis 
calculations. The steeply sloping lines on the far left are a simple
extrapolation of a PIB model with $n = 0$.  This extrapolation is 
disrupted as indicated for the three models.
At the nucleosynthesis scale, three
possible fluctuation spectra are considered as described in the text.}
\label{fig4}
\end{figure}

\end{document}